\documentclass[10pt,twocolumn,a4paper]{article}

\usepackage[left=20mm, right=15mm, top=15mm, bottom=15mm]{geometry}

\usepackage{subfig}
\usepackage{flushend}
\usepackage{indentfirst}
\usepackage{graphics}
\usepackage{amsmath}
\usepackage{graphicx}
\usepackage{epstopdf}
\usepackage{float}

\usepackage[font=footnotesize,labelfont=bf]{caption}
\usepackage[labelsep=period]{caption}

\graphicspath{{./figure/}}

\begin{document}

\title{\huge \textbf{A Chiellini type integrability condition for the generalized first
kind Abel  differential equation}}

\date{}

\twocolumn[
\begin{@twocolumnfalse}
\maketitle

\author{\textbf{Tiberiu Harko}$^{1,*}$, \textbf{Francisco S. N. Lobo}$^{2}$, \textbf{M. K. Mak}$^{3}$\\\\
\footnotesize $^{1}${Department of Mathematics, University College London, Gower Street, London
WC1E 6BT, United Kingdom}\\
\footnotesize $^{2}$Centro de Astronomia e Astrof\'{\i}sica da Universidade de Lisboa, Campo
Grande, Ed. C8 1749-016 Lisboa, Portugal\\
\footnotesize $^{3}$Department of Computing and Information Management, Hong Kong
Institute of Vocational Education, Chai Wan, Hong Kong, P. R. China\\
\footnotesize $^{*}$Corresponding Author: t.harko@ucl.ac.uk}\\\\\\

\end{@twocolumnfalse}
]

\noindent \textbf{\large{Abstract}} \hspace{2pt} The Chiellini integrability condition of the first order first kind Abel equation $%
dy/dx=f(x)y^2+g(x)y^3$ is extended to the case of the general Abel equation
of the form $dy/dx=a(x)+b(x)y+f(x)y^{\alpha -1}+g(x)y^{\alpha }$, where $\alpha \in \Re$, and $%
\alpha > 1$. In the case $\alpha =2$ the generalized Abel equations reduces to a Riccati
type equation, for which a Chiellini type integrability condition is
obtained.\\


\noindent\hrulefill

\section{Introduction}

The third degree polynomial, first kind first order Abel differential
equation \cite{Pol}
\begin{equation}
\frac{dy}{dx}=f(x)y^{2}+g(x)y^{3},  \label{1}
\end{equation}%
where the coefficients $f(x)$ and $g(x)$ are real functions of the variable $%
x$, plays an important role in many physical and mathematical problems. For
example, the Li\'{e}nard type second order differential equation,
given by \cite{Lien}%
\begin{equation}
\frac{d^{2}y}{dx^{2}}+f\left( y\right) \frac{dy}{dx}+g\left( y\right) =0,
\label{LL}
\end{equation}
 can be trasformed to the Li\'{e}nard system,
represented by the autonomous system of differential equations
\begin{equation}
\frac{dy}{dx}=u,\qquad \frac{du}{dx}=-f(y)u-g(y),
\end{equation}
describing nonlinear damped oscillations of a dynamical system. With the help of the substitution $u=1/v$, the
Li\'{e}nard system can easily be reduced to a first-order first kind Abel
differential equation of the form
\begin{equation}
\frac{dv}{dy}=f(y)v^{2}+g(y)v^{3}.
\end{equation}

An integrability condition for the Abel Eq.~(\ref{1}) was obtained by
Chiellini \cite{Chiel, kamke}, which can be formulated as follows \cite{Chiel}:

\textbf{Chiellini integrability condition.} A first kind Abel type
differential equation of the form given by Eq.~(\ref{1}) can be exactly
integrated if the functions $f(x)$ and $g(x)$ satisfy the condition
\begin{equation}
\frac{d}{dx}\frac{g(x)}{f(x)}=kf(x), \qquad k = \mathrm{constant}, \qquad k\neq 0.
\end{equation}

This integrability result for the Abel equation has been applied for
obtaining exact solutions of second order differential equations that can be
reduced to an Abel type equation in \cite{B1} - \cite{Mak4}.

The Abel Eq.~(\ref{1}) is the ``reduced'' form of the general first kind Abel equation $%
dy/dx=a(x)+b(x)y+f(x)y^{2}+g(x)y^{3}$, where $a(x),b(x),f(x),g(x)$ are
arbitrary real functions. A possibility of generalizing the reduced Abel Eq. (\ref%
{1}) is to consider that the non-linear terms in $y$ (the term $y^{2}$ and $%
y^{3}$, respectively), appearing in the right hand side of the general Abel
equation are at some arbitrary powers $y^{\alpha -1}$ and $y^{\alpha }$, respectively,
where $\alpha >1$ is a real number. We call this equation a generalized first
kind Abel equation.

It is the purpose of the present paper to extend the Chiellini integrability
condition to the generalized Abel equations of the first kind. If the
coefficients of the equation satisfy two integrability conditions, the
general solution can be found through quadratures. In the particular case $%
\alpha =2$, the generalized Abel equations reduces to a Riccati type equation, and
a Chiellini type integrability condition for the Riccati equation is
obtained.

The present paper is organized as follows. The integrability condition for
the generalized Abel equation is obtained in Section~\ref{sect2}. The
Chiellini type integrability condition for the Riccati equation is obtained
in Section~\ref{sect3}. We conclude our results in Section~\ref{sect4}.

\section{The Chiellini integrability condition for generalized Abel equations%
}

\label{sect2}

We introduce a generalization of the Abel first kind differential Eq.~(\ref%
{1}) through the following:

\textbf{Definition.} The first order nonlinear ordinary differential
equation, which can be given as
\begin{equation}
\frac{dy}{dx}=a(x)+b(x)y+f(x)y^{\alpha -1}+g(x)y^{\alpha }, \quad \alpha \in \Re, \quad \alpha >1, \label{gen}
\end{equation}%
where $a(x),b(x),f(x),g(x)\in C^{\infty }(I)$ are arbitrary functions
defined on a real interval $I\subseteq \Re $, $a(x),b(x),f(x),g(x)\neq
0,\forall x\in I $, and $\alpha \in \Re$ satisfies the condition $\alpha >1$, is called
the generalized first kind Abel type differential equation.

By introducing a new function $p(x)$, defined as
\begin{equation}
y(x)=e^{\int {b(x)dx}}p(x),
\end{equation}%
Eq.~(\ref{gen}) becomes
\begin{eqnarray}
\frac{dp(x)}{dx}&=&a(x)e^{-\int {b(x)dx}}+f(x)e^{\left( \alpha -2\right) \int {b(x)dx}%
}p^{\alpha -1}(x)\nonumber\\
&&+g(x)e^{\left( \alpha -1\right) \int {b(x)dx}}p^{\alpha }(x).  \label{gen1}
\end{eqnarray}%
We assume now that the functions $b(x)$, $f(x)$ and $g(x)$ satisfy the
generalized Chiellini condition
\begin{eqnarray}
\frac{d}{dx}\left[ \frac{f(x)}{g(x)e^{\int {b(x)dx}}}\right] =-k_1\frac{%
f^{\alpha }(x)}{g^{\alpha -1}\left( x\right) }e^{-\int {b(x)dx}}, \nonumber\\
k_1={\rm constant}, \qquad k_1\neq 0, \label{cond1}
\end{eqnarray}%
where $k_1\in \Re$, $k_1\neq 0$,  is an arbitrary constant. Then, by introducing the
transformation
\begin{equation}
p(x)=\frac{f(x)}{g(x)e^{\int {b(x)dx}}}s(x),
\end{equation}%
Eq.~(\ref{gen1}) becomes
\begin{equation}
\frac{ds}{dx}=\frac{g(x)}{f(x)}a(x)+\frac{f^{\alpha -1}(x)}{g^{\alpha -2}(x)}\left(
s^{\alpha }+s^{\alpha -1}+k_1s\right) .  \label{s}
\end{equation}

Therefore we have obtained the following generalization of the Chiellini
integrability condition:

\textbf{Theorem 1.} If the coefficients $a(x)$, $b(x)$, $f(x)$ and $g(x)$ of
the generalized first kind Abel Eq.~(\ref{gen}) satisfy the conditions (\ref{cond1})
and
\begin{eqnarray}
a(x)=k_{2}\frac{f^{\alpha }(x)}{g^{\alpha -1}(x)}=-\frac{k_{2}}{k_{1}}e^{\int {b(x)dx}}%
\frac{d}{dx}\left[ \frac{f(x)}{g(x)e^{\int {b(x)dx}}}\right] ,
 \label{a}
\end{eqnarray}%
respectively, where $k_{2}\in \Re $, $k_2\neq 0$, is an arbitrary constant, the general
solution of Eq.~(\ref{gen}) is given by
\begin{equation}
y(x)=\frac{f(x)}{g(x)}s(x),  \label{12}
\end{equation}%
where $s(x)$ is a solution of the equation
\begin{equation}
\left| \frac{g(x)e^{\int b\left( x\right) dx}}{f(x)}\right| =K^{-1}e^{F\left[
s(x),k_{1},k_{2}\right] },  \label{13}
\end{equation}%
$K\neq 0$ is an arbitrary integration constant, and
\begin{equation}
F\left[ s(x),k_{1},k_{2}\right] =k_{1}\int {\frac{ds}{%
s^{\alpha }+s^{\alpha -1}+k_{1}s+k_{2}}}.  \label{14}
\end{equation}

In order to obtain the above Theorem we have used the result, which represents the
generalization of the transformations introduced for the third degree Abel
polynomial equation in \cite{Rosu1},
\begin{eqnarray}
&&\int\frac{f^{\alpha -1}(x)}{g^{\alpha -2}(x)}dx= \nonumber\\
&&-\frac{1}{k_1}\int \frac{1}{%
f(x)/g(x)e^{\int b\left( x\right) dx}}d\left[ \frac{f(x)}{g(x)e^{\int
b\left( x\right) dx}}\right] =\nonumber\\
&&\frac{1}{k_1}\ln \left| \frac{g(x)e^{\int
b\left( x\right) dx}}{f(x)}\right| +K_{0},  \label{k}
\end{eqnarray}
where $K_{0}\neq 0$ is an arbitrary constant of integration, which is related to
the constant $K$ by the relation $K=\exp \left(k_1K_0\right)$.

The generalized Chiellini integrability condition, given by Eq.~(\ref{cond1}%
), can be reformulated as a Bernoulli type differential equation for $f(x)$,
\begin{equation}
\frac{df}{dx}=\left[ \frac{1}{g(x)}\frac{dg(x)}{dx}+b(x)\right] f(x)-k_{1}%
\frac{f^{\alpha }(x)}{g^{\alpha -2}(x)},
\end{equation}%
with the general solution given by
\begin{eqnarray}
f(x)&=&g(x)e^{\int {b(x)dx}}\Bigg[ K_{1}-(1-\alpha )k_{1}\times \nonumber\\
&&\int {e^{(\alpha -1)\int{%
b\left( x\right) dx}}g(x)dx}\Bigg] ^{1/(1-\alpha)},
\alpha \neq 1, \nonumber\\ \label{condi1}
\end{eqnarray}%
where $K_{1}$ is an arbitrary constant of integration.

As a differential equation for $g(x)$, the integrability condition Eq.~(\ref%
{cond1}) can be reformulated as
\begin{equation}
g^{\alpha -3}(x)\frac{dg}{dx}=\Bigg[ \frac{1}{f(x)}\frac{df(x)}{dx}-b(x)\Bigg]
g^{\alpha -2}(x)+k_{1}f^{\alpha -1}(x),
\end{equation}%
with the general solution given by
\begin{eqnarray}
g(x)&=&f(x)e^{-\int {b(x)dx}}\Bigg[ K_{2}+(\alpha -2)k_{1}\times \nonumber\\
&&\int {f(x)e^{(\alpha -2)\int {%
b\left( x\right) dx}}dx}\Bigg] ^{1/(\alpha -2)}, \qquad  \alpha \neq
2, \nonumber\\ \label{condi2}
\end{eqnarray}%
where $K_{2}$ is an arbitrary constant of integration.

Therefore we have obtained the following:

\textbf{Theorem 2}. If the coefficients $a(x)$, $b(x)$, $f(x)$ and $g(x)$ of
the generalized second type Abel differential Eq.~(\ref{gen}) satisfy the
conditions (\ref{condi1}), or (\ref{condi2}), and (\ref{a}), then the
differential equation is exactly integrable, and its general solution can be
obtained from Eqs.~(\ref{12})-(\ref{14}).

Theorem 2 represents the generalization to the case of the generalized Abel
equation of the results obtained initially for the third degree Abel
differential equation in \cite{Rosu1}, and further discussed in \cite{Mak4}.

A particular case of the integrability of the generalized Abel equation can
be obtained by assuming that the constants $k_{1}$ and $k_{2}$ satisfy the
condition $k_{1}=k_{2}$. Then the function $F\left[ s(x),k_{1},k_{2}\right] $
can be written as
\begin{equation}
F\left[ s(x),k_{1},k_{2}\right] =k_{1}\int {\frac{ds}{\left(
s^{\alpha -1}+k_{1}\right) \left( s+1\right) }}.
\end{equation}

In many cases the above integral can be obtained exactly. As a particular
case we consider the value $\alpha =3/2$. Hence the generalized Abel type equation
takes the form
\begin{equation}
\frac{dy}{dx}=a(x)+b(x)y+f(x)y^{1/2}+g(x)y^{3/2}.  \label{ex}
\end{equation}%
Therefore if the coefficients $a(x)$, $b(x)$, $f(x)$ and $g(x)$ of the
generalized Abel differential Eq.~(\ref{ex}) satisfy the conditions
\begin{equation}
\frac{d}{dx}\left[ \frac{f(x)}{g(x)e^{\int {b(x)dx}}}\right] =-k_{1}\frac{%
f^{3/2}(x)}{g^{1/2}\left( x\right) }e^{-\int {b(x)dx}},
\end{equation}
and
\begin{equation}
a(x)=k_{1}\frac{%
f^{3/2}(x)}{g^{1/2}(x)},
\end{equation}%
respectively, then the solution of Eq.~(\ref{ex}) is given by
\begin{equation}
y(x)=\frac{f(x)}{g(x)}s(x),
\end{equation}
where $s(x)$ is a solution of the equation
\begin{eqnarray}  \label{24}
&&K\left| \frac{g(x)e^{\int b\left( x\right) dx}}{f(x)}\right| =e^{k_{1}\int
\frac{ds}{\left( \sqrt{s}+k_{1}\right) \left( s+1\right) }}=\nonumber\\
&&\frac{\left(
1+s\right) ^{k_{1}^{2}/\left( 1+k_{1}^{2}\right) }}{\left( k_{1}+\sqrt{s}%
\right) ^{2k_{1}^{2}/\left( 1+k_{1}^{2}\right) }}e^{\left[ 2k_{1}/\left(
1+k_{1}^{2}\right) \right] \arctan \sqrt{s}}.
\end{eqnarray}

By power expanding the right hand side of Eq.~(\ref{24}) one can easily
obtain some approximate solutions of Eq.~(\ref{ex}).

\section{The generalized Chiellini integrability condition for the Riccati
equation}

\label{sect3}

In the case $\alpha =2$ the generalized second type Abel differential Eq.~(\ref%
{gen}) reduces to a Riccati type equation \cite{Pol} of the form
\begin{equation}
\frac{dy}{dx}=a(x)+\left[ b(x)+f(x)\right] y+g(x)y^{2}.  \label{ric}
\end{equation}

Therefore the generalized Chiellini condition can be extended to the case of
Eq.~(\ref{ric}), giving the following integrability condition for the
Riccati equation:

\textbf{Theorem 3.} If the coefficients $a(x)$, $f(x)$, $b(x)$ and $g(x)$ of
the Riccati Eq. (\ref{ric}) satisfy the conditions
\begin{eqnarray}
&&\frac{d}{dx}\left[ \frac{f(x)}{g(x)e^{\int {b(x)dx}}}\right] =-k_{1}\frac{%
f^{2}(x)}{g\left( x\right) }e^{-\int {b(x)dx}}, \nonumber\\
&&k_1={\rm constant}, \qquad k_1\neq 0, \label{cond2}
\end{eqnarray}%
and
\begin{equation}
a(x)=k_{2}\frac{f^{2}(x)}{g(x)}, \qquad k_2={\rm constant}, \qquad  k_2\neq 0, \label{a1}
\end{equation}%
respectively, where $k_{1},k_{2}\in \Re $, $k_1\neq 0$, $k_2\neq 0$,  are arbitrary constants, then the
general solution of the Riccati Eq.~(\ref{ric}) is given by
\begin{eqnarray}
y(x)&=&\frac{f(x)}{g(x)}\Bigg\{ \frac{\sqrt{-\Delta }}{2}\tan \left[ \frac{%
\sqrt{-\Delta }}{2k_{1}}\ln \left| \frac{Kg(x)e^{\int {b(x)dx}}}{f(x)}%
\right| \right] \nonumber\\
&&-\frac{1+k_{1}}{2}\Bigg\} ,\qquad \Delta <0,  \label{28}
\end{eqnarray}%
\begin{eqnarray}
y(x)&=&-\frac{f(x)}{g(x)}\left\{ \frac{k_{1}}{\ln  \left| Kg(x)e^{\int {%
b(x)dx}}/f(x)\right|  }+\frac{1+k_{1}}{2}\right\} , \nonumber\\
&&\Delta =0,
\label{29}
\end{eqnarray}%
\begin{eqnarray}
y(x)&=&-\frac{f(x)}{g(x)}\Bigg\{ \frac{\sqrt{\Delta }}{2}\tanh \left[ \frac{%
\sqrt{\Delta }}{2k_{1}}\ln \left| \frac{Kg(x)e^{\int {b(x)dx}}}{f(x)}\right| %
\right] \nonumber\\
&&+\frac{1+k_{1}}{2}\Bigg\} , \Delta >0,  \label{30}
\end{eqnarray}%
where $K$ is an arbitrary integration constant, and $\Delta =\left( 1+k_{1}\right) ^{2}-4k_{2}$.

The integrability condition of the Riccati Eq.~(\ref{cond2}) can be written
as a differential equation for $f(x)$,
\begin{equation}
\frac{df(x)}{dx}=\left[ \frac{1}{g(x)}\frac{dg(x)}{dx}+b(x)\right]
f(x)-k_{1}f^{2}(x),
\end{equation}%
with the general solution given by
\begin{equation}
f(x)=\frac{g(x)e^{\int {b(x)dx}}}{K_{3}+k_{1}\int {g(x)\exp \left[ \int {%
b\left( x\right) dx}\right] dx}},  \label{condi3}
\end{equation}%
where $K_{3}$ is an arbitrary integration constant. As a differential
equation for $g(x)$, the Riccati equation integrability condition (\ref%
{cond2}) can be formulated as
\begin{equation}
\frac{dg(x)}{dx}=\left[ \frac{1}{f(x)}\frac{df(x)}{dx}-b(x)+k_{1}f(x)\right]
g(x),
\end{equation}%
\begin{equation}
g(x)=K_{4}f(x)e^{-\int {\left[ b(x)-k_{1}f(x)\right] dx}},  \label{condi4}
\end{equation}%
where $K_{4}$ is an arbitrary integration constant.

Therefore we have obtained the following:

\textbf{Theorem 4}. If the coefficients $a(x)$, $f(x)$, $b(x)$ and $g(x)$ of
the Riccati equation (\ref{ric}) satisfy the conditions (\ref{condi3}), or (%
\ref{condi4}), and (\ref{a1}), then the Riccati equation is exactly
integrable, and its general solution is given by Eqs.~(\ref{28})-(\ref{30}).

\section{Concluding remarks}

\label{sect4}

In the present paper, we have obtained an exact integrability condition for
the generalized first kind Abel equation. The general solution of the generalized
nonlinear Abel differential equation can be obtained by quadratures
if the four coefficients of the equation satisfy two consistency conditions.
The constraint imposes severe restrictions, limiting the number of possible
solutions that can be obtained in this way. An integrability condition for
the Riccati equation, representing a particular case of the generalized Abel
equation, was also obtained. Some applications of the present formalism to
the Abel equations appearing in physical problems will be presented in a
future paper.


\begin{thebibliography}{99}
\bibitem{Pol} A. D. Polyanin and V. F. Zaitsev, Handbook of Exact Solutions
for Ordinary Differential Equations, Chapman \& Hall/CRC, Boca Raton,
London, New York, Washington, D. C. (2003).

\bibitem{Lien} A. Li\'{e}nard, Revue g\'{e}n\'{e}rale de l'\'{e}lectricit%
\'{e} \textbf{23}, 901-912, and 946-954 (1928).

\bibitem{Chiel} A. Chiellini, Sull'integrazione dell'equazione differenziale $y'+Py^2+Qy^3=0$,  Bollettino dell' Unione Matematica Italiana,
\textbf{10}, 301-307 (1931).

\bibitem{kamke} E. Kamke, Differentialgleichungen:  L\"{o}sungsmethoden und L\"{o}sungen, Chelsea, New York, (1959).

\bibitem{B1} I. Bandi$\acute{\mathrm{c}}$, Sur le crit$\grave{\mathrm{e}}$re
d' int$\acute{\mathrm{e}}$grabilit$\acute{\mathrm{e}}$ de l' $\acute{\mathrm{%
e}}$quation diff$\acute{\mathrm{e}}$rentielle g$\acute{\mathrm{e}}$n$\acute{%
\mathrm{e}}$ralis$\acute{\mathrm{e}}$e de Li$\acute{\mathrm{e}}$nard,
Bollettino dell' Unione Matematica Italiana, \textbf{16}, 59 (1961).

\bibitem{B2} I. Bandi$\acute{\mathrm{c}}$, Sur les $\acute{\mathrm{e}}$%
quations diff$\acute{\mathrm{e}}$rentielles non-lin$\acute{\mathrm{e}}$aires
quasihomog$\grave{\mathrm{e}}$nes $\grave{\mathrm{a}}$ deux dimensions du
premier et du deuxi$\grave{\mathrm{e}}$me ordres, Bollettino dell' Unione
Matematica Italiana, \textbf{17}, 8191 (1962).

\bibitem{Mak1} M. K. Mak, H. W. Chan, and T. Harko, Solutions Generating
Technique for Abel-Type Nonlinear Ordinary Differential Equations, Comput.
Math. Appl. \textbf{41}, 1395-1401 (2001).

\bibitem{Mak2} M. K. Mak and T. Harko, New Method for Generating General
Solution of Abel Differential Equation, Comput. Math. Appl. \textbf{43},
91-94 (2002).

\bibitem{Mak3} T. Harko, M. K. Mak, Relativistic Dissipative Cosmological
Models and Abel Differential Equation, Comput. Math. Appl. \textbf{46},
849-853 (2003).

\bibitem{Rosu1} S. C. Mancas and H. C. Rosu, Integrable dissipative
nonlinear second order differential equations via factorizations and Abel
equations,  Phys. Lett. {\bf A 377}, 1234 - 1238 (2013); arXiv:1212.3636.

\bibitem{Rosu2} S. C. Mancas and H. C. Rosu, Integrable Ermakov-Pinney equations with nonlinear Chiellini "damping", arXiv:1301.3567 (2013).

\bibitem{Mak4} T. Harko, F. S. N. Lobo, and M. K. Mak, A class of exact
solutions of the Li\'{e}nard type ordinary non-linear differential equation,
arXiv:1302.0836 (2013).
\end{thebibliography}
\end{document}